\documentclass[reprint, amsmath,amssymb, aps, superscriptaddress, floatfix,longbibliography]{revtex4-1}

\usepackage[utf8]{inputenc}
\usepackage{graphicx}
\usepackage{dcolumn}
\usepackage{bm}
\usepackage{braket}
\usepackage{color}
\usepackage{units}
\usepackage{dsfont}
\usepackage{float}
\usepackage{hyperref}
\usepackage{xcolor}

\usepackage{tabularx}

\begin{document}

\title{Photonic quantum walk with ultrafast time-bin encoding}

\author{Kate L. Fenwick}
\affiliation{National Research Council of Canada, 100 Sussex Drive, Ottawa, Ontario K1A 0R6, Canada}
\affiliation{Department of Physics, University of Ottawa, Advanced Research Complex, 25 Templeton Street, Ottawa ON Canada, K1N 6N5}
\author{Fr\'ed\'eric Bouchard}
\email{frederic.bouchard@nrc-cnrc.gc.ca}
\affiliation{National Research Council of Canada, 100 Sussex Drive, Ottawa, Ontario K1A 0R6, Canada}
\author{Duncan England}
\affiliation{National Research Council of Canada, 100 Sussex Drive, Ottawa, Ontario K1A 0R6, Canada}
\author{Philip J. Bustard}
\affiliation{National Research Council of Canada, 100 Sussex Drive, Ottawa, Ontario K1A 0R6, Canada}
\author{Khabat Heshami}
\affiliation{National Research Council of Canada, 100 Sussex Drive, Ottawa, Ontario K1A 0R6, Canada}
\affiliation{Department of Physics, University of Ottawa, Advanced Research Complex, 25 Templeton Street, Ottawa ON Canada, K1N 6N5}
\author{Benjamin Sussman}
\affiliation{National Research Council of Canada, 100 Sussex Drive, Ottawa, Ontario K1A 0R6, Canada}
\affiliation{Department of Physics, University of Ottawa, Advanced Research Complex, 25 Templeton Street, Ottawa ON Canada, K1N 6N5}

\begin{abstract}
The quantum walk (QW) has proven to be a valuable testbed for fundamental inquiries in quantum technology applications such as quantum simulation and quantum search algorithms. Many benefits have been found by exploring implementations of QWs in various physical systems, including photonic platforms. Here, we propose a novel platform to perform quantum walks using an ultrafast time-bin encoding (UTBE) scheme. This platform supports the scalability of quantum walks to a large number of steps while retaining a significant degree of programmability. More importantly, ultrafast time bins are encoded at the picosecond time scale, far away from mechanical fluctuations. This enables the scalability of our platform to many modes while preserving excellent interferometric phase stability over extremely long periods of time without requiring active phase stabilization. Our 18-step QW is shown to preserve interferometric phase stability over a period of 50 hours, with an overall walk fidelity maintained above $95\%$.
\end{abstract}

\maketitle

Random processes are fundamental in nature and ubiquitous in life. The random walk (RW), one of the most important stochastic processes in probability theory, provides a means of emulating the evolution of random processes~\cite{masuda2017random}. The quantum mechanical analog of the classical RW is the quantum walk (QW)~\cite{aharonov1993quantum,kempe2003quantum}. Unlike its classical counterpart, the QW is described by a quantum mechanical wave function which spreads across the different possible outcomes. Quantum interference effects in the wave function lead to drastically different outcomes than those observed in a classical RW, making the QW a powerful tool for observing the manifestation of non-classical effects. Beyond its fundamental importance in observing quantum mechanical wave function evolution, the QW has become an invaluable tool with applications in quantum simulations~\cite{strauch2006relativistic,witthaut2010quantum,lee2015quantum}, quantum search algorithms~\cite{shenvi2003quantum,childs2004spatial,qu2022deterministic}, quantum transport~\cite{mulken2011continuous}, universal quantum computations~\cite{childs2009universal}, and machine learning~\cite{flamini2023reinforcement}. Various other physical systems have been used to perform QWs, such as atoms~\cite{karski2009quantum}, trapped ions~\cite{schmitz2009quantum,zahringer2010realization}, nuclear magnetic resonance~\cite{ryan2005experimental}, and superconducting qubits~\cite{gong2021quantum}, but, due to the interference present in QWs, photonic systems lend themselves naturally to the task.

Experimental realizations of photonic QWs have been reported with walkers in both the spatial and temporal degrees of freedom using different implementations, such as bulk optics~\cite{do2005experimental,bouwmeester1999optical,broome2010discrete,kitagawa2012observation,cardano2015quantum,cardano2016statistical,cardano2017detection,nejadsattari2019experimental,geraldi2019experimental,d2020two,esposito2022quantum,di2023ultra}, fiber loops~\cite{schreiber2010photons,schreiber20122d,jeong2013experimental,lorz2019photonic,geraldi2021transient,bagrets2021probing,held2022driven,pegoraro2023dynamic}, fiber cavities~\cite{boutari2016large}, integrated photonics~\cite{peruzzo2010quantum,sansoni2012two,harris2017quantum,pitsios2017photonic}, and measurement-based approaches~\cite{de2022measurement}. Nevertheless, practical limitations to the number of steps, the overall phase stability, the loss per step, and the degree of control over individual QW steps, pose challenges that could be overcome by using ultrafast time-bin encoding (UTBE). In the UTBE scheme, information is encoded in photon arrival time, with femtosecond-duration pulses spaced by several picoseconds. The generation, manipulation, and detection of ultrafast time bins can be achieved using ultrashort pulses and birefringent crystals~\cite{kupchak2017time,bouchard2022quantum,donohue2013coherent}. Time bin encoding~\cite{brendel1999pulsed} generally offers good phase stability in transmission because of common path propagation; however, inter-bin manipulations usually require multiple optical elements with active phase stabilization. In contrast, UTBE can manipulate time bins in a common path, using a single element with excellent passive phase stability and low loss per step.

Readout of ultrafast time bins is unattainable by most single photon detectors; however, it is easily achieved by optical gating. One particularly promising method of optical gating uses an all-optical Kerr gate ~\cite{england2021perspectives}, which has previously been demonstrated to achieve near-unit efficiency gating on picosecond timescales~\cite{kupchak2019terahertz,bouchard2021achieving}. The Kerr gate has also been implemented at sub-picosecond timescales, on orders of hundreds-of-femtoseconds~\cite{fenwick2020carving}. In this article, we propose and experimentally demonstrate an UTBE scheme for QWs using birefringent $\alpha$-barium borate ($\alpha$-BBO) crystals, which is read-out by an all-optical ultrafast Kerr gate. We measure the step-by-step evolution of the QW for single-photon input states with varied polarization and time-bin distributions. The QW is demonstrated to achieve high fidelity for up to 18 steps and remarkably high stability over long periods of time. The inherent phase stability of our UTBE scheme paves the way for scaling up experimental QW implementations to larger numbers of steps while maintaining low loss at each step. Moreover, each step can be addressed separately, providing greater control over the QW parameters and potential for step-by-step programmability.

\begin{figure*}[t!]
	\centering
		\includegraphics[width=1\textwidth]{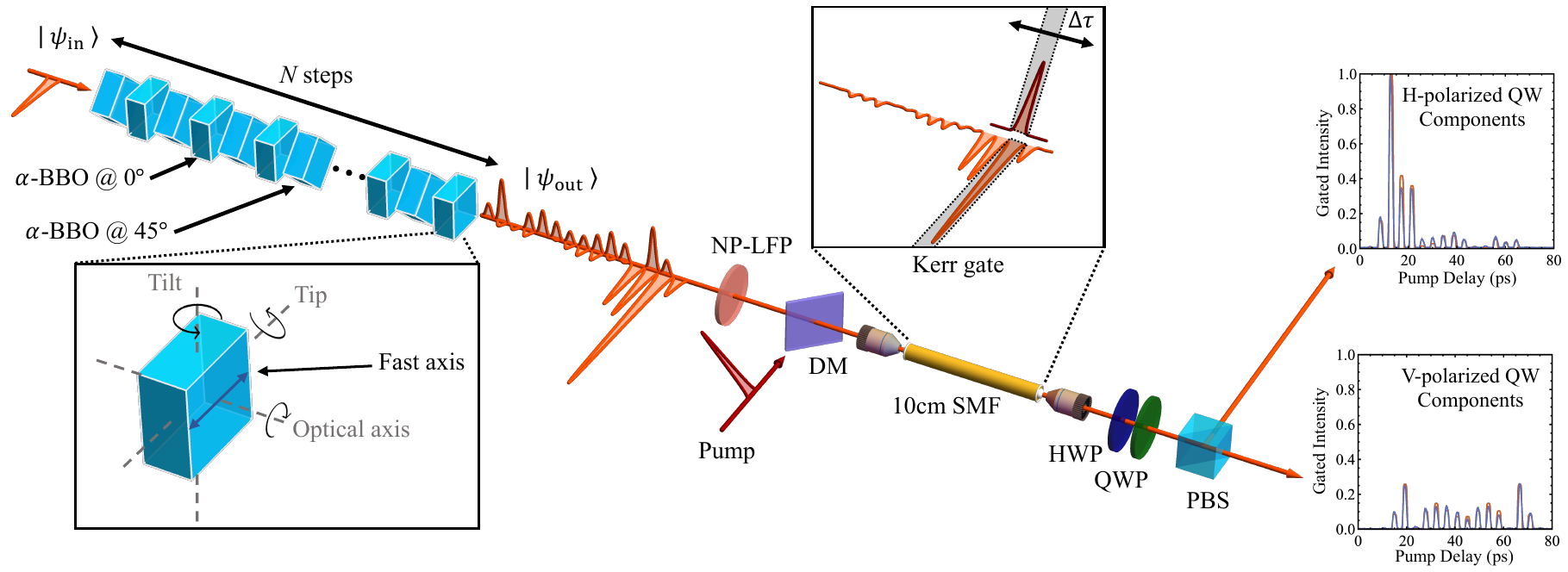}
	\caption{\textbf{Conceptual experimental setup.} Simplified experimental setup of the ultrafast time-bin QW. The input state, $|\psi_{\text{in}} \rangle$, propagates through an $N$-step QW. Each step in the QW consists of a birefringent $\alpha$-BBO crystal, which has three rotational degrees of freedom (shown in the bottom inset). The output state, $|\psi_{\text{out}} \rangle$, is sampled by an optical Kerr gate, where a strong pump pulse induces polarization rotation only where it overlaps with $|\psi_{\text{out}} \rangle$ (shown in the top inset). Prior to recombination with the pump on a dichroic mirror (DM), the QW output state is sent through a nanoparticle linear film polarizer (NP-LFP), which can be rotated to measure either its H- or V-polarized components. Scanning the pump, $\Delta\tau$, allows for readout of the entire QW output state. Plots on the right show experimental data (blue) collected for the H- and V-polarized QW components, compared with model (red). PBS: polarizing beamsplitter, HWP: half-waveplate, QWP: quarter-waveplate, SMF: single mode fiber. Further experimental details are provided in Appendix A.}
	\label{fig:concept}
\end{figure*}

A QW acts on a quantum particle represented by a tensor product of two different basis states. One basis will define the space of the ``walker'' and the other basis will define the space of the ``coin''. As the quantum particle ``walks through'' each step of the QW, its probability distribution in the walker space is changed depending on its input coin state. Here, we use a single-photon-level pulse as the quantum particle in the photonic QW. Ultrafast time bins, with basis states ${|t_i\rangle \in \{ |t_0\rangle, |t_1\rangle, |t_2\rangle, ... \}}$, define the space of the walker. The polarization state of the photon takes the coin's role, with basis states ${|H\rangle}$ and ${|V\rangle}$, corresponding to horizontally and vertically polarized light, respectively. 

\begin{figure*}[t!]
	\centering
		\includegraphics[width=1\textwidth]{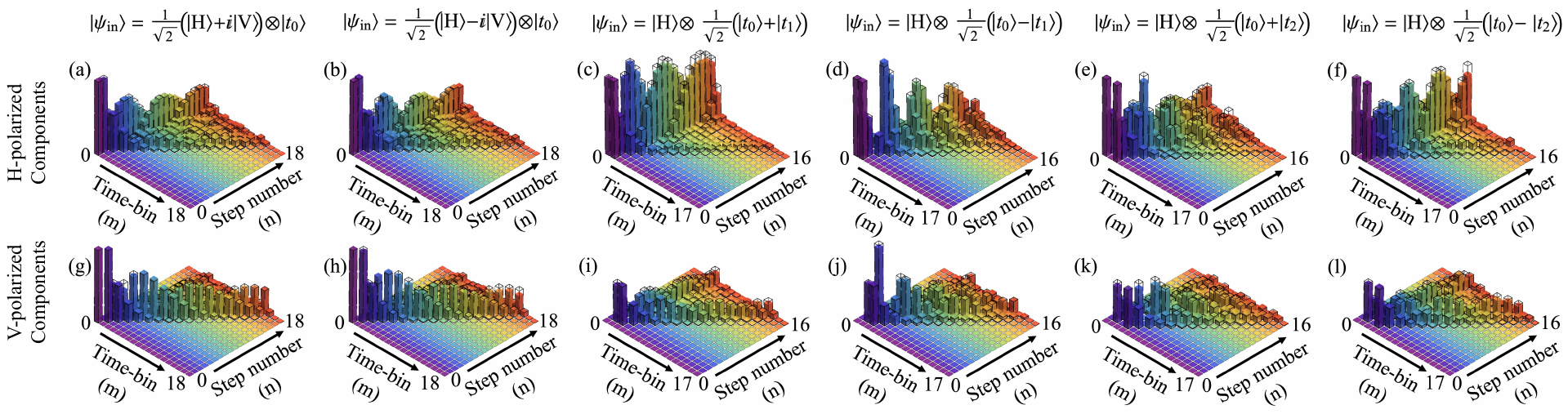}
	\caption{\textbf{Discretized step-by-step evolution of the QW for different input states.} Each column shows the results obtained for a different input state (defined at the top of each column). The top row (a-f) shows the H-polarized components of the output state. The bottom row (g-l) shows the V-polarized components of the output state. The color indicates the step-by-step evolution (purple to red) of the experimental QW data, which is compared with the model (black outlines). Here, we can observe the emerging two-peaked distribution expected from quantum interference in the QW~\cite{kempe2003quantum}. When the input state is polarized along one of the $\alpha$-BBO crystal axes, such as $| \psi_{\text{in}} \rangle = | H \rangle \otimes 1/\sqrt{2}(| t_{\text{0}} \rangle + | t_{\text{1}} \rangle)$ or $| \psi_{\text{in}} \rangle = | H \rangle \otimes 1/\sqrt{2}(| t_{\text{0}} \rangle - | t_{\text{2}} \rangle)$, we observe an asymmetry in the two-peaked distribution. When the input state is in a superposition of these two polarizations, such as $| \psi_{\text{in}} \rangle = 1/\sqrt{2}(| H \rangle\pm i| V \rangle) \otimes | t_{\text{0}} \rangle$, the two-peaked output distribution will be symmetric. For the input states $| \psi_{\text{in}} \rangle = | H \rangle \otimes 1/\sqrt{2}(| t_{\text{0}} \rangle - | t_{\text{1}} \rangle)$ and $| \psi_{\text{in}} \rangle = | V \rangle \otimes 1/\sqrt{2}(| t_{\text{0}} \rangle + | t_{\text{2}} \rangle)$, we do not observe as strong a splitting of the distribution due to the phase of the input state. Details on data collection and processing are provided in Appendix D.}
	\label{fig:3Devolutions}
\end{figure*}

A single step, $n$, of our photonic QW, is given by the unitary step operator,
\begin{equation}
\hat{U}_n = \hat{S}_n \cdot  \hat{C}_n. 
 \label{eq:step}
\end{equation}
The shift operator, $\hat{S}_n$, raises or maintains the time-bin state of the single photon depending on its polarization state,
\begin{equation}
\hat{S}_n = \sum_m \Big(|H\rangle \langle H| \otimes |t_{m}\rangle \langle t_m| + |V\rangle \langle V| \otimes |t_{m+1}\rangle \langle t_m| \Big), 
 \label{eq:shift}
\end{equation}
and the quantum mechanical analog of the coin flip is given by the coin operator, $\hat{C}_n$, acting on the coin subspace such that
\begin{align}
    \hat{C}_n &= \sum_m \Big( \cos\Big(\frac{\Omega_n}{2}\Big) |H\rangle \langle H| + e^{i \gamma_n} \sin\Big(\frac{\Omega_n}{2}\Big) |H\rangle \langle V| \nonumber \\
    & \hspace{0.3cm}+ e^{-i \gamma_n} \sin\Big(\frac{\Omega_n}{2}\Big) |V\rangle \langle H| - \cos\Big(\frac{\Omega_n}{2}\Big) |V\rangle \langle V| \Big) \nonumber \\
    & \hspace{6.0cm} \otimes |t_m\rangle \langle t_m |  \nonumber \\
\end{align}
where $\Omega_n$ and $\gamma_n$ are parameters that can be adjusted to tune the probability amplitude and phase of the state, respectively. Different types of walks can be realized by varying the parameters of the coin operator, for example biased~\cite{vstefavnak2009recurrence} and step-dependent coin~\cite{xue2015experimental} walks. 

To achieve the step operator in Eq.~\ref{eq:step}, we use a birefringent crystal, $\alpha$-BBO, with a very low loss per step of $< - 0.045$\,dB. The thickness of the $\alpha$-BBO crystals sets the temporal separation between time-bins. In this work, we use $\alpha$-BBO crystals each with a thickness of 10\,mm, providing a temporal separation of 4.3\,ps at the signal wavelength. The orientation of each $\alpha$-BBO crystal sets the coin operator, $\hat{C}_n$. As seen in Fig.~\ref{fig:concept}, each step in the QW we present here is individually programmable, consisting of a single $\alpha$-BBO crystal with access to three rotational degrees of freedom. In this work, we set $\Omega_n=\pi/2$ by rotating the fast axis of each $\alpha$-BBO to alternating values of $45^{\circ}$ and $0^{\circ}$, for odd and even steps, respectively. We also tune the tip and tilt of each $\alpha$-BBO to set the phase to $\gamma_n = 0$. 

The QW comprises a line of $N$ birefringent $\alpha$-BBO crystals, each of which performs one step in the walk corresponding to the operator $\hat{U}_n$. The total evolution of an $N$-step QW is therefore given by
\begin{equation}
|\psi_\mathrm{out}^{(N)} \rangle = \prod_{n=1}^{N} \hat{U}_n|\psi_{\text{in}} \rangle.
\end{equation}
In general, the single photon input state can be written as
\begin{equation}
    |\psi_{\text{in}} \rangle = \sum_m \Big( \alpha_m|H  \rangle + \beta_m|V  \rangle \Big) \otimes a_m|t_m \rangle, 
\end{equation}
where $\alpha_m$, $\beta_m$, and $a_m$ are complex numbers with $\sum_m (|\alpha_m|^2+|\beta_m|^2) |a_m|^2=1$. Control over the input state is also necessary in order to program the QW for quantum simulations. The input state can be prepared in different polarization states, using quarter- and half-waveplates (QWPs and HWPs), and across multiple time bins using an appropriate sequence of $\alpha$-BBO crystals.

The result of the QW is obtained by measuring the probability distribution of $|\psi_\mathrm{out}^{(N)} \rangle$ for each polarization component. As these time bins occur on ultrafast timescales, an all-optical Kerr gate is necessary to measure the output state. The Kerr gate relies on cross-phase modulation via the optical Kerr effect, a third-order ($\chi^{(3)}$) optical nonlinearity where a strong pulse of light generates a localized intensity-dependent refractive index modulation in the medium through which it propagates. By doing so, the strong pulse induces a birefringence that can rotate the polarization of a secondary optical signal. In this work, the Kerr gate medium is a short (10\,cm) piece of single-mode fiber (SMF). Depicted in Fig.~\ref{fig:concept}, when crossed-polarizers are placed along the signal beam path before and after the SMF, the signal that is temporally-overlapped with the pump will undergo polarization rotation. The entire QW output state is measured by scanning pump delay, $\Delta\tau$, and detecting Kerr-gated intensity. We achieve a Kerr-gate efficiency of $>99\%$, for which a mathematical description is provided in Appendix B. An overall system efficiency is outlined in Appendix C.

Changing the input state of the QW changes its evolution. We thus examine the step-by-step evolution of the QW for various input states, shown in Fig.~\ref{fig:3Devolutions}, in order to demonstrate the robustness of our protocol. A PBS, HWP, and QWP are used to generate different input polarization states. Input states with more than one time bin can be prepared using the $\alpha$-BBO crystals from the first steps of the QW. For the results presented here, single time-bin input state evolution is shown up to 18 steps. For the two time-bin input state evolution, the first two $\alpha$-BBO crystals in the QW are removed and repurposed as needed for input state preparation, so a 16-step QW is performed.

\begin{figure}[t!]
	\centering
		\includegraphics[width=0.5\textwidth]{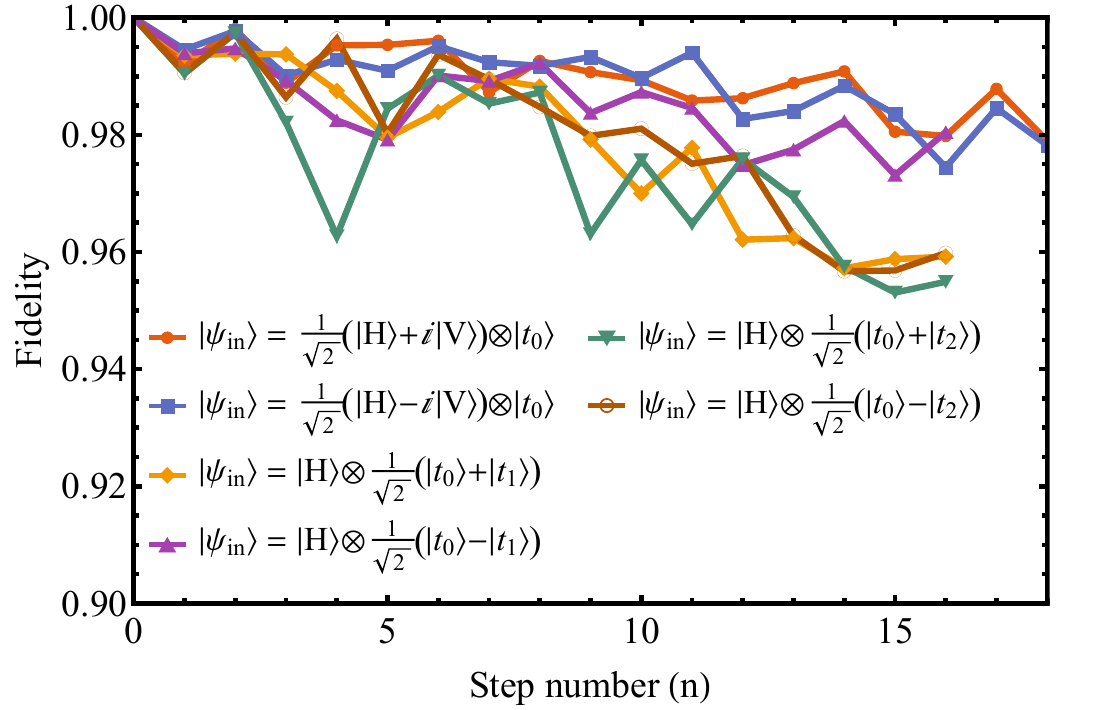}
	\caption{\textbf{Step-by-step fidelity of the QW for different input states.} Calculated fidelity after each step for the six different input states measured. Fidelities remain $>97\%$ and $>95\%$ over the entire evolution of each 18-step QW and 16-step QW, respectively.}
	\label{fig:fidelityBoth}
\end{figure}

The measured evolution compares well to theoretical evolution, which is indicated by the transparent black outlines. The agreement between experiment and theory is quantified by a fidelity calculation
\begin{equation}
    F_n(X,Y) = \Big(\sum_{i=0}^{n}\sum_{j=1}^2 \sqrt{p_{i,j} q_{i,j}} \Big)^2,
\end{equation}
where $F_n(X,Y)$ describes the overlap between the experimental state, $X$, and theoretical state, $Y$, for a given number of steps, $n$, in the QW. The index $i$ sums over the time bins and runs from the input time bin, $i=0$, to the total number of time bins present after $n$ steps. The index $j$ sums over the two polarization components (H and V). The probabilities of the experimental and theoretical QW outputs of being in a given polarization and time-bin are denoted by $p_{i,j}$ and $q_{i,j}$, respectively. The fidelity calculated for the six different input states is shown in Fig.~\ref{fig:fidelityBoth}. The experimental data are in good agreement with the theory, with fidelities remaining above $95\%$ for all input states. 

\begin{figure}[b!]
	\centering
		\includegraphics[width=0.5\textwidth]{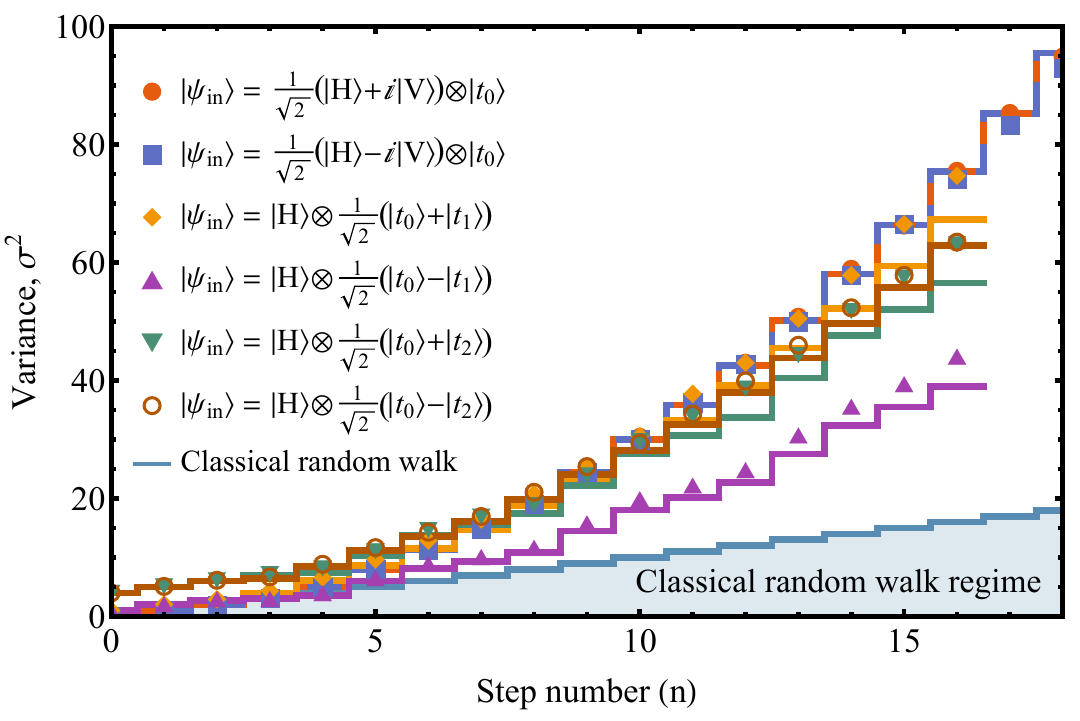}
	\caption{\textbf{Step-by-step variance of the QW}. Experimental QW variances (data points) are compared with their respective theoretical variances (step-like lines) for various input states. The variance of a classical RW (for the case of a single classical particle that begins the walk in a single “time-bin”), is shaded at the bottom (light blue) for reference. For some input states, experimental values are higher than theoretical values due to background noise across all time bins in the measurement.}
	\label{fig:variance}
\end{figure}

A key feature of the QW is that its probability distribution spreads considerably faster than that of a classical random walk. This can lead to exponentially faster hitting times when traversing a graph~\cite{shenvi2003quantum}, an extremely useful attribute for quantum search algorithms. The standard deviation or variance of the QW output can be calculated to show this feature~\cite{broome2010discrete}. We calculate the step-by-step variance of our QW, shown in Fig.~\ref{fig:variance}, for the different input states investigated in this work. We observe that the QW variance grows exponentially with step number, as opposed to the linear growth in variance expected for a classical random walk.

\begin{figure}[t!]
	\centering
		\includegraphics[width=0.5\textwidth]{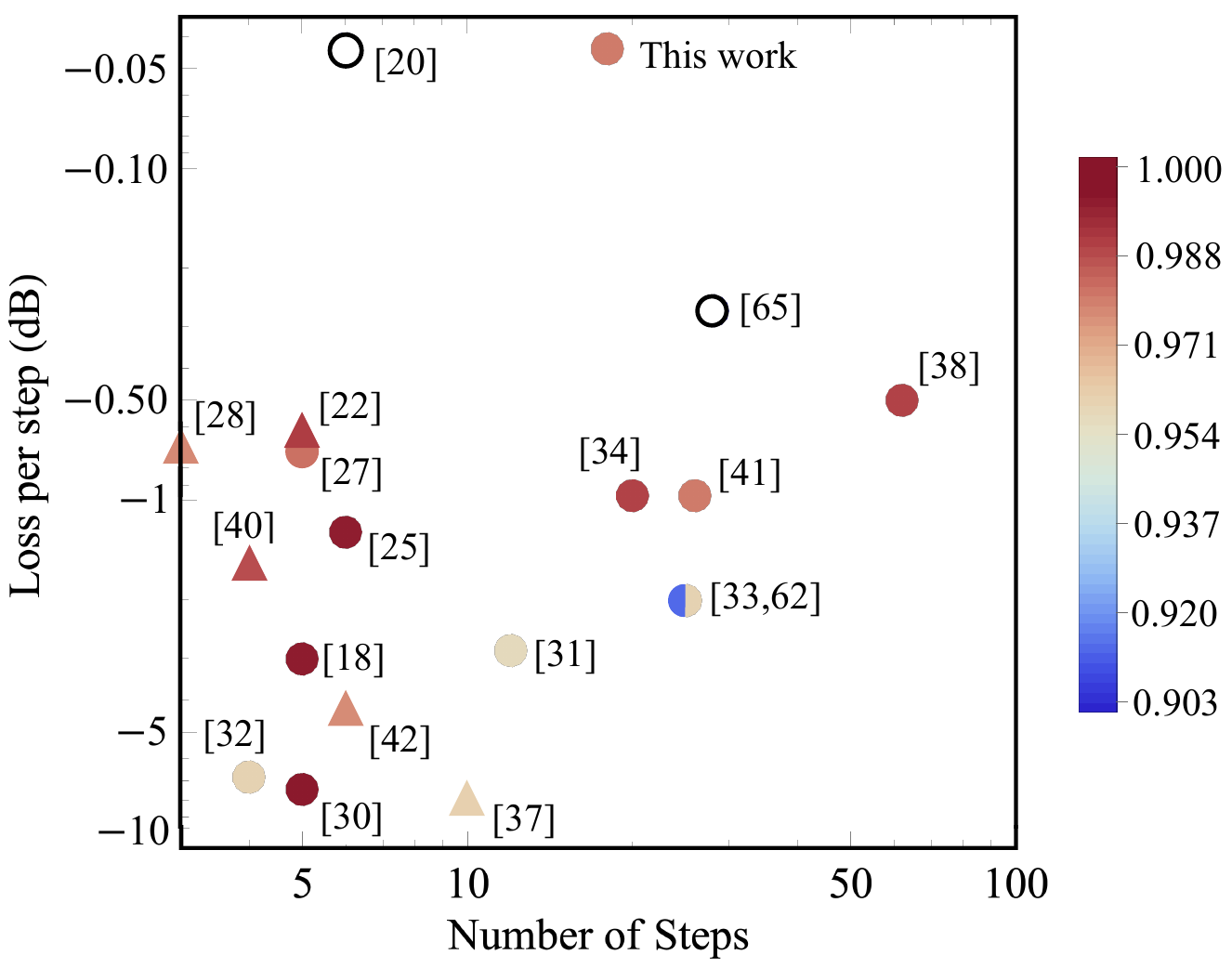}
	\caption{\textbf{Current QW Landscape}. The loss per step is plotted against the number of steps for the experimental discrete photonic QWs that are cited. The color of each point corresponds to the reported fidelity, a scale for which is shown to the right of the plot. Circles and triangles correspond to QWs with one-photon and two-photon inputs, respectively. For references in which fidelity was not reported, the data points are white with black outlines. Note that this plot does not constitute an exhaustive list of all QW studies in the literature, and work that did not report on all three attributes (loss per step, number of steps, and fidelity) were unable to be included. See Table~\ref{tab:1} in Appendix F for a more comprehensive list. Here, we include our 18-step walk with 98\% fidelity.}
	\label{fig:literatureQWs}
\end{figure}

Not only does our QW protocol demonstrate high fidelity for a large number of steps, it is also capable of preserving the interferometric phase stability for prolonged periods. We verified the stability over a 50-hour period (see Appendix E). With this high stability, and a low loss of $< -0.045$\,dB per step, our QW platform is expected to be easily scalable to a larger number of steps. To visualize where our platform fits into the current QW landscape with these attributes, it is compared to a sample of other platforms reported in the literature, on the basis of number of steps achieved, loss per step, and overall fidelity in Fig.~\ref{fig:literatureQWs}. The sample of references included does not constitute an exhaustive list, but does provide an overall picture of the current experimental capabilities of discrete photonic QWs.

The combined robustness, programmability, and stability of our QW platform make it a particularly promising solution for scalability to a larger number of output modes while remaining space-efficient. These attributes are crucial for realizing QWs in regimes where classical verification is not possible or debilitatingly slow, for example efficiently simulating Hamiltonian evolution~\cite{low2017optimal,childs2003exponential}. QWs have already been demonstrated for simulation purposes such as in topological investigations~\cite{kitagawa2012observation, cardano2016statistical, cardano2017detection}. In these cases, fewer steps than we report here were required. It is anticipated that our QW could be extended into even more complex regimes in a variety of ways, depending on the problem at hand. Increasing the number of steps in the QW, thereby allowing access to more modes, is one option. This can be done by simply adding more $\alpha$-BBO crystals to the QW signal beam path. With only 31 steps it is expected that the QW could be used in simulating parton showers~\cite{bepari2022quantum}, cascades of radiation produced in high-energy particle collisions. Another feasible near-term option requires 40 linear QW steps for the simulation of symmetries, topological phases, and bound states~\cite{asboth2012symmetries}. 
QW platforms can also borrow inspiration from experiments which use two-dimensional arrays of quantum particles for quantum simulation, such as Rydberg atom arrays~\cite{karski2009quantum,kalinowski2022bulk} or trapped ion arrays~\cite{schmitz2009quantum,zahringer2010realization,macdonell2022predicting}. Extending the space of the QW to two dimensions~\cite{tang2018experimental,oliveira2006decoherence,schreiber20122d} can be used to simulate quantum systems with a natural mapping to two dimensions, such as in topological systems~\cite{chen2018observation}. Beyond this, some QW schemes have even been demonstrated up to four-dimensions \cite{hamilton2011quantum,lorz2019photonic}. It would be straightforward to increase the dimensionality of the ultrafast time-bin QW platform, where multiple time-bin bases could be achieved using birefringent crystals of different thickness.

Since our QW platform is programmable at each step, it is realistic to consider the implications of implementing a time-dependent coin operator, where the coin depends on the temporal variable~\cite{banuls2006quantum,schreiber2011decoherence}. Time-dependent coin operators have been used to demonstrate periodic revivals of a QW distribution \cite{xue2015experimental}, and can demonstrate emergence of a Parrondo paradox in QWs without resorting to intricate high-dimensional coins~\cite{pires2020parrondo}. A time-dependent coin could be implemented in our ultrafast time-bin QW by adding a Kerr gate in the QW itself. Step-dependent coin operation could also be achieved, without active Kerr gates, by simply rotating and tilting the $\alpha$-BBO crystals. These capabilities would allow for walking along any arbitrary graph, thereby extending the QW protocol for use in quantum search algorithms~\cite{shenvi2003quantum,childs2004spatial,qu2022deterministic,feder2006perfect}. One could even envision implementing machine learning to program the QW~\cite{mathew2021raspberry,hinrichs2020neural}, or using the QW for machine learning~\cite{flamini2023reinforcement}. 

Although increasing the number of steps or dimensionality of the QW may be adequate in some cases, multi-photon input states are required if one is to simulate more complex, multi-particle quantum dynamics. Two-photon QWs have previously been demonstrated in various platforms~\cite{defienne2016two, owens2011two, peruzzo2010quantum}, and have even been used to reveal previously-unknown two-photon effects~\cite{simon2020quantum} and perform two-photon tomography~\cite{titchener2016two}. Multi-photon entangled states are extremely sensitive to interferometric noise, as is demonstrated by the N00N state~\cite{afek2010high} and other highly-entangled multi-photon states~\cite{guo2020distributed}. A crucial requirement in the scaling of QWs to multi-photon states is, therefore, maintaining interferometric stability across the entire QW over extended periods of time. The prolonged phase stability of our QW platform makes it a strong candidate for this task. Extra caution must be taken with multi-photon input states, since particle losses can drastically alter the output~\cite{pegoraro2023dynamic}; however the high efficiency of our scheme makes it realistic to move to multi-photon input states. In order to measure $n$-fold coincidences, where $n$ is the number of input photons, more Kerr gates would be required. It is possible to implement multiple Kerr-gates for such a task~\cite{bouchard2023measuring}; however, another solution could be to implement fast detectors, such as superconducting nanowire single-photon detectors (SNSPDs), which can achieve timing jitter as low as 3\,ps~\cite{korzh2020demonstration}. In this case, thicker $\alpha$-BBO crystals could be used to achieve ultrafast time-bins with larger temporal separation. It should be noted that measuring $n$-fold coincidences with SNSPDs is not trivial and would require multiplexing across $\sim n^2$ fast detectors. Furthermore, careful consideration of detector dead time would be required. 

Our proposed UTBE scheme for discrete photonic QW demonstrates the key ingredients for scalability, including high interferometric phase stability for a large number of modes, programmability at each step, and low loss per step. With these combined capabilities we offer a pathway to scale programmable QWs with simple experimental tools, opening up new avenues for exploring problems that are intractable by classical computers. 

\section*{Acknowledgments}
The authors acknowledge that the National Research Council of Canada (NRC) headquarters is located on the traditional unceded territory of the Algonquin Anishinaabe and Mohawk people. This work is supported by the Vanier Canada Graduate Scholarships (Vanier CGS) program. We thank Aaron Goldberg, Andrew Proppe, Brayden Freitas, Denis Guay, Doug Moffatt, Guillaume Thekkadath, Kent Bonsma-Fisher, Rune Lausten, and Yingwen Zhang for their support and insightful discussions.

\section*{Appendix A: Experimental details}
\setcounter{equation}{0}
\let\oldtheequation\theequation
\renewcommand{\theequation}{A\oldtheequation}

\begin{figure*}[t!]
	\centering
		\includegraphics[width=1\textwidth]{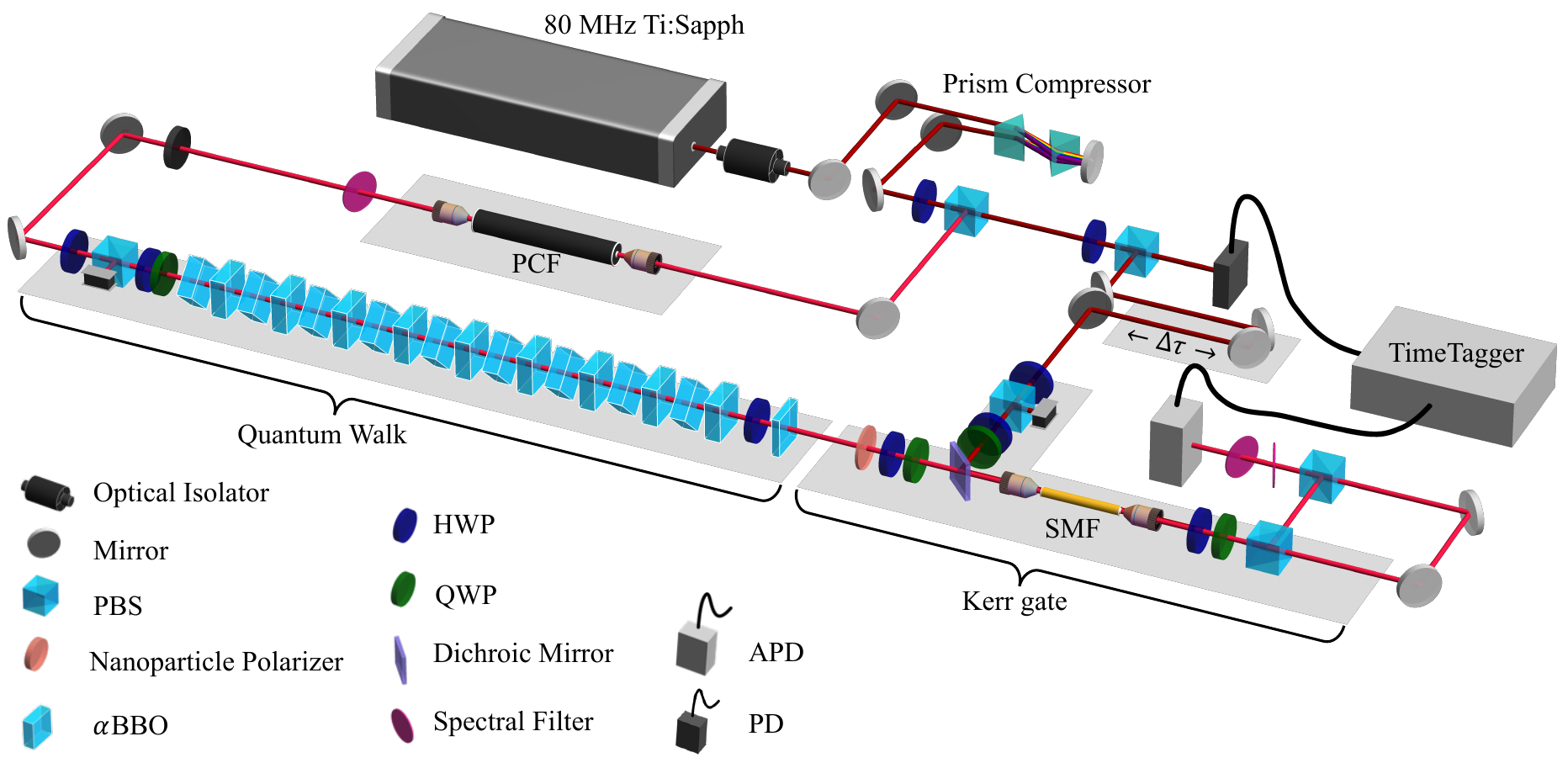}
	\caption{\textbf{Experimental Setup.} Detailed experimental setup of the ultrafast time-bin QW. In the above configuration, all 18 $\alpha$-BBO crystals are used in the QW (\textit{i.e.,} none are used for input state generation). PBS: polarizing beamsplitter, HWP: half-waveplate, QWP: quarter-waveplate, APD: avalanche photodiode, PD: photodiode, SMF: single-mode fiber; PCF: photonic crystal fiber.}
	\label{fig:experiment}
\end{figure*}
The full experimental setup in this work is shown in Fig.~\ref{fig:experiment}. Both the signal and pump originate from an 80\,MHz titanium sapphire (Ti:Sapph) laser, centered at a wavelength of $\lambda_\mathrm{pump}=800~\mathrm{nm}$ with a spectral bandwidth of $\Delta\lambda_\mathrm{pump}=11~\mathrm{nm}$. The output pulses from the Ti:Sapph laser are first sent through an optical isolator to avoid unwanted backreflected light re-entering the laser cavity. After the optical isolator, a prism compressor is used for dispersion compensation. The pulses are then split into two paths (one for the signal and one for the pump) by a HWP and a polarizing beamsplitter (PBS), which provide control over the pulse energy in the two paths.

Supercontinuum generation (SCG) in a 12\,cm photonic crystal fiber (PCF) generates signal pulses~\cite{dudley2006supercontinuum}. When pumped with pulses from the Ti:sapph laser, the PCF (NKT Photonics, FemtoWHITE CARS) generates a dual peak spectrum which is tunable with input pulse energy. As the pump pulse energy is increased, the two SCG peaks will shift further away from the pump central wavelength and increase in bandwidth. In the sub-nJ pump pulse energy regime, SCG in the PCF provides access to a signal wavelength range of 600 to 1100\,nm. Input pulses with energy $\sim0.25$\,nJ generate a spectrum with one of the dual peaks centered at 720\,nm, which is spectrally-filtered to $\Delta \lambda_\mathrm{signal}=5~\mathrm{nm}$ using a band-pass filter. Neutral density (ND) filters are then used to attenuate the signal pulses to a mean photon number of $\mu \approx 0.8$ photons per pulse. The input photon polarization state is prepared using a PBS, QWP, and a HWP, after which it enters the QW. It should be noted that, for the configuration shown in Fig.~\ref{fig:experiment}, the signal pulses are prepared in a single time bin, $| \psi_\mathrm{in} \rangle = ( \alpha|H  \rangle + \beta|V  \rangle ) \otimes |t_0\rangle$. 

Input states with more than one time bin can be prepared using the $\alpha$-BBO crystals from the first steps of the QW. Here, we prepare two time-bin states given by
\begin{equation}
    |\psi_{\text{in}} \rangle = |H  \rangle \otimes  \frac{1}{\sqrt{2}} \Big( |t_0 \rangle + e^{i\nu}|t_k \rangle  \Big),
\end{equation}
where the phase, $\nu$, and the arrival of the $t_k$ time bin are varied. The position, $k$, of the $t_k$ time bin depends on the number of $\alpha$-BBO crystals used to generate the input state, whereas the tip and tilt of the crystals sets $\nu$. In the data presented here, $k=1,2$ and $\nu=0,\pi$. The first two $\alpha$-BBO crystals in the QW are removed and repurposed as needed for input state preparation, so a 16-step QW is performed. Note that for all input states, a PBS is added after the crystal(s) used for input state generation, to set the input polarization state to $| H \rangle$. 

The $\alpha$-BBO crystals comprising the QW are housed in kinematic rotation mounts, providing control over all three rotational degrees of freedom. By controlling each crystal's rotational degree of freedom, $\hat{C}_n$ can be set arbitrarily for each step of the QW. Alignment of the entire QW is done in a cascading manner, where each crystal is tuned relative to the previous one. 

The QW output is recombined with the pump pulses on a dichroic mirror, so that the Kerr-gated measurement can be made. Prior to recombination, the pump is prepared as followed. The pump pulse energy is first tuned, using a HWP and PBS combination, to maximize the gating efficiency. A PBS, QWP, and HWP are used to prepare the pump polarization so that it is $45^\circ$ relative to the signal polarization at the 10\,cm SMF input to maximize gate efficiency. Finally the pump pulses are sent through a variable delay line, $\Delta\tau$, to delay the pump relative to the output of the QW. Scanning $\Delta\tau$ allows for temporal traces of the QW signal to be read-out by the Kerr gate at the picosecond timescale. The pump and signal are both coupled to the 10\,cm SMF with $80\%$ efficiency. 

The QW signal is Kerr-gated, with $>99\%$ internal efficiency, by the pump pulse within the 10\,cm SMF (Thorlabs, S630-HP). After the Kerr gate, a QWP and HWP compensate for any intrinsic polarization rotation or birefringence within the 10\,cm SMF. A PBS then splits the gated and non-gated QW signal along two different optical path lengths (separated on the nanosecond timescale). Spectral filters attenuate the pump and any noise generated by the pump inside the 10\,cm SMF. The gated and non-gated signals are both detected by an avalanche photodiode (APD) and processed using a time-to-digital converter.

\section*{Appendix B: Kerr Gate Efficiency}
\setcounter{equation}{0}
\renewcommand{\theequation}{B\oldtheequation}

The Kerr gate efficiency, $\eta$, is given by~\cite{kanbara1994highly}
\begin{equation}
    \eta(T) = \text{sin}^2(2\theta) \text{sin}^2\Bigg(\frac{\Delta\phi(T)}{2}\Bigg), 
\end{equation}
where $\theta$ is the angle between the signal and pump polarization, and $\Delta\phi(T)$ is the induced nonlinear birefringence in the SMF due to the presence of the pump pulse. A unit gating efficiency is achieved when $\theta=\pi/4$ and $\Delta \phi = \pi$. The nonlinear phase shift is given by~\cite{agrawal2001applications}
\begin{equation}
    \Delta\phi(T) = \frac{8\pi n_2}{3\lambda_{\text{s}}}\int_0^L I_{\text{p}}(T-d_w z)dz, 
\end{equation}
where $n_2$ is the nonlinear refractive index of the SMF, $\lambda_{\text{s}}$ is the signal wavelength, and $z$ is the propagation distance within the SMF of length $L$. As the SMF is a dispersive medium, the pump and signal will experience temporal walkoff, which is defined by $d_w=v_{gp}^{-1} - v_{gs}^{-1}$, where $v_{gp}$ and $v_{gs}$ are the group velocities of the pump and signal pulses, respectively. The intensity profile of the pump, $I_p$ is expressed in a frame that moves with the signal~\cite{kupchak2019terahertz}, $T=t-z/v_{gs}$, where $t$ is the time in the laboratory frame.

\section*{Appendix C: Overall System Efficiency}

The efficiency of the overall QW setup is defined by each optical element it comprises. Here, we quote the loss, in dB, for the different optical elements used. Silver mirrors: -0.088\,dB; HWP: -0.223\,dB; QWP: -0.223\,dB; nanoparticle thin-film polarizer (NP-TFP): -0.044\,dB; telescope: -0.094\,dB; dichroic mirror: -0.044\,dB; Kerr gate (includes aspheric lenses and 10\,cm SMF): -1.192\,dB; PBS: -0.706\,dB; spectral filters: -1.871\,dB; 1\,m SMF to detector: -0.706\,dB; APD: -2.218\,dB. The majority of the $\alpha$-BBO crystals used in the QW have an AR coating which leads to -0.044\,dB loss (Newlight Photonics, ABT5100-A-P, a-cut); however, some crystals have an AR coating which leads to -0.269\,dB loss (Newlight Photonics, ABT5100-A-AR750, a-cut). Combining the various experimental losses (including those of optical elements not depicted in Fig.~\ref{fig:experiment}), we find an overall system efficiency of $\sim 5\%$.
\newline

\section*{Appendix D: Data collection and processing}

\begin{figure*}[t!]
	\centering
		\includegraphics[width=0.96\textwidth]{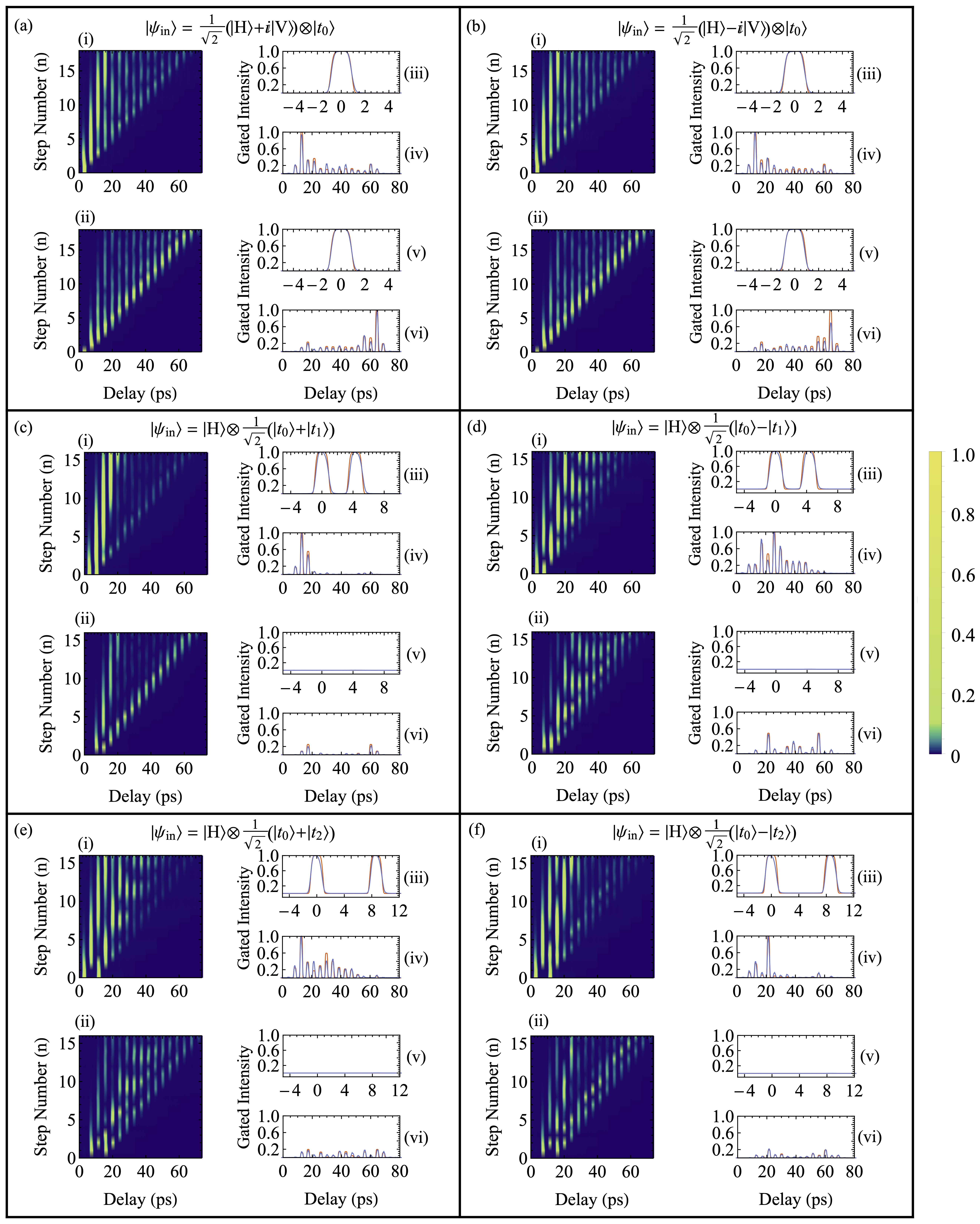}
	\caption{\textbf{Measured data for the UTBE QW for various input states.} Boxes (a)-(f) correspond to various input states (indicated at the top of each box). In each box, the step-by-step evolution of the H- and V-polarized components of the QW are shown in (i) and (ii), respectively. A color scale for these is shown to the right of all the plots. Each box compares the experimental traces (blue) with the theoretical traces (red) for: the H-polarized input state in (iii), the H-polarized output state in (iv), the V-polarized input state in (v), and the V-polarized output state in (vi). Experimental traces agree well with theoretical traces.}
    \label{fig:traces}
\end{figure*}

The QW output state is sampled by the Kerr gate after each step is added. Fig.~\ref{fig:traces} shows the temporal traces of the six different input states (see subsets (iii) for H-polarized components and (v) for V-polarized components) and their corresponding outputs (see subsets (iv) for H-polarized components and (vi) for V-polarized components) after the 18- or 16-step QWs. The input and output experimental traces, shown in blue, compare well to the theory, shown in red. The experimental step-by-step evolution of the QW for each input state is also shown in Fig.~\ref{fig:traces} (see subsets (i) for H-polarized components and (ii) for V-polarized components). We note here that the time bin, $|t_0\rangle$, occurs at a delay of 0\,ps. After each step of the QW, one new time bin is added to the output state at a delay of $4.3n$\,ps, where $n$ is the step number. At the last step in the QW, the time bins span a range of $0$ to $4.3N$\,ps, where $N$ is the total number of steps. This leads to a step-by-step evolution that expands along the delay axis as the QW evolves. The QW output traces are discretized by taking the peak value at each time-bin delay, $\Delta\tau=4.3n$, where $n$ is the step number.

\clearpage

\section*{Appendix E: Interferometric phase stability}

\begin{figure}[h!]
	\centering
		\includegraphics[width=0.5\textwidth]{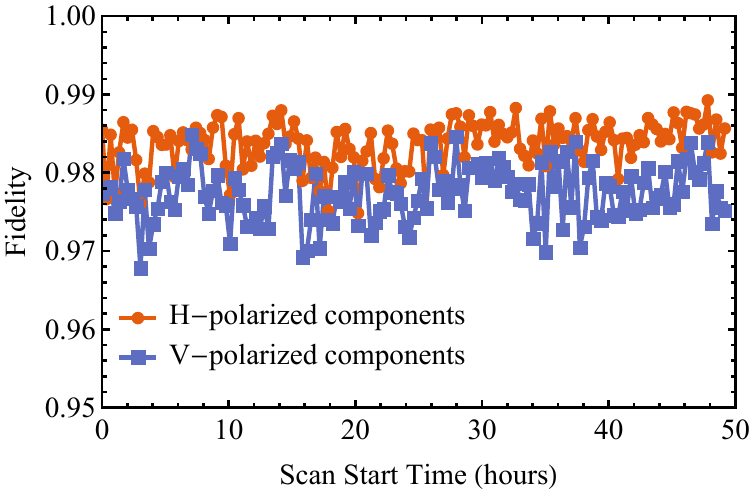}
	\caption{\textbf{Interferometric phase stability of UTBE QW}. Calculated fidelity for the H- and V-polarized components of an 18-step QW with an input state of: $|\psi_{\text{in}}\rangle = |H\rangle \otimes |t_0\rangle$. Note that the overall fidelity combines both the H- and V-polarized components, but here we opt to show that they both have interferometric phase stability. The fidelity of the V-polarized components is slightly less than that of the H-polarized components, as there is less V-polarized light for this configuration. The data presented here was collected continuously over 50 hours to showcase the stability of the QW platform.}
	\label{fig:stability}
\end{figure}

The advantage of using ultrafast time bins is highlighted in Fig.~\ref{fig:stability}, where we see high stability for the 19-mode interferometric output associated with our 18-step QW over a 50 hour continuous measurement window. 

\section*{Appendix F: QW landscape}

A summary of experimental implementations of discrete-time photonic QWs is provided in Table~\ref{tab:1}. Although efforts were made to ensure a comprehensive list, it is not exhaustive. We observe that our QW platform is capable of achieving a larger number of steps than many other bulk optics implementations. Integrated photonics and fiber/optical loop schematics achieve a large number of steps, but they have a higher loss per step. Finally, we note how the 320-~\cite{di2023ultra} and 400-step~\cite{de2022measurement} QWs differ from other implementations. In ~\cite{di2023ultra}, 320-step QWs are simulated by a single ``step'': 3 liquid-crystal metasurfaces, encoded with the appropriate unitary. In ~\cite{de2022measurement}, measurements on rotated bases implement the targeted unitary.

\begin{table*}[t]
  \centering
\begin{tabular}{p{0.7cm} p{1.2cm} p{1.5cm} p{2.0cm} p{1.7cm} p{1.5cm} p{2cm}}
 \hline\hline
 Year & Reference & Number of steps & Loss per step [dB] & Overall fidelity & Number of photons & Platform \\ [0.5ex] 
 \hline\hline
 2005 & \cite{do2005experimental} & 5 & -3.010 & 0.9984 & 1 photon & Bulk optics \\ [0.5ex]

 2008 & \cite{ribeiro2008quantum} & 1 & -- & -- & 1 photon & Bulk optics \\ [0.5ex]

 2010 & \cite{schreiber2010photons} & 5 & -7.447 & 0.999 & 1 photon & Fiber loop\\ [0.5ex]

 2010 & \cite{broome2010discrete} & 6 & -0.044 & Distance: 0.031 & 1 photon & Bulk optics\\ [0.5ex]

 2011 & \cite{hamilton2011quantum} & 12 & -- & -- & 1 photon & Optical loop\\ [0.5ex]
  
 2011 & \cite{schreiber2011decoherence} & 28 & -0.269 & Distance: 0.052 & 1 photon & Fiber loop\\ [0.5ex]

 2012 & \cite{kitagawa2012observation} & 7 & -- & -- & 1 photon & Bulk optics\\ [0.5ex]

 2012 & \cite{sansoni2012two} & 4 & -1.600 & 0.988 & 2 photons & Integrated photonics\\ [0.5ex]

 2012 & \cite{schreiber20122d} & 12 & -2.840 & 0.957 & 1 photon & Fiber network\\ [0.5ex]

 2013 & \cite{crespi2013anderson} & 8 & -- & 0.951 & 2 photons & Integrated photonics\\ [0.5ex]

 2013 & \cite{jeong2013experimental} & 4 & -6.840 & 0.96 & 1 photon & Optical loop\\ [0.5ex]

 2014 & \cite{grafe2014chip} & 16 & -- & 0.918 & 1 photon & Integrated photonics\\ [0.5ex]

 2015 & \cite{cardano2015quantum} & 5 & -0.655 & 0.991 & 2 photons & Orbital angular momentum\\ [0.5ex]

 2015 & \cite{xue2015experimental} & 16 & -- & 0.81 & 1 photon & Bulk optics\\ [0.5ex]

 2016 & \cite{boutari2016large} & 62 & -0.500 & 0.99 & 1 photon & Optical ring cavities\\ [0.5ex]

 2016 & \cite{cardano2016statistical} & 6 & -- & -- & 1 photon & Orbital angular momentum\\ [0.5ex]

 2017 & \cite{cardano2017detection} & 7 & -- & -- & 1 photon & Orbital angular momentum\\ [0.5ex]

 2017 & \cite{pitsios2017photonic} & 6 & -4.437 & 0.977 & 2 photons & Integrated photonics\\ [0.5ex]

 2017 & \cite{harris2017quantum} & 26 & -0.969 & 0.98 & 1 photon & Integrated photonics\\ [0.5ex]

 2018 & \cite{chen2018observation} & 25 & -2.007 & 0.96 & 1 photon & Optical loop\\ [0.5ex]

 2019 & \cite{nejadsattari2019experimental} & 6 & -1.249 & 0.9982 & 1 photon & Orbital angular momentum\\ [0.5ex]

 2019 & \cite{geraldi2019experimental} & 7 & -- & 0.999 & 2 photons & Bulk optics\\ [0.5ex]

 2019 & \cite{lorz2019photonic} & 25 & -2.007 & 0.912 & 1 photon & Optical loop\\ [0.5ex]

 2020 & \cite{d2020two} & 5 & -0.706 & 0.982 & 1 photon & Transverse momentum\\ [0.5ex]

 2021 & \cite{geraldi2021transient} & 20 & -0.969 & 0.99 & 1 photon & Optical loop\\ [0.5ex]

 2022 & \cite{qu2022deterministic} & 12 & -- & 0.992 & 1 photon & Bulk optics\\ [0.5ex]

 2022 & \cite{esposito2022quantum} & 3 & -0.706 & 0.9773 & 2 photons & Transverse momentum\\ [0.5ex]

 2022 & \cite{de2022measurement} & 400 & -- & 0.98 & 1 photon & Measurement-based\\ [0.5ex]

 2023 & \cite{di2023ultra} & 320 & -0.706 & 0.77 & 1 photon & Liquid-crystal metasurfaces\\ [0.5ex]

 2023 & \cite{pegoraro2023dynamic} & 10 & -8.239 & 0.961 & 2 photons & Optical loop\\ [0.5ex]

 2023 & This work & 18 & -0.044 & 0.98 & 1 photon & Bulk optics (UTBE)\\ [0.5ex]
 \hline
\end{tabular}
  \caption{\textbf{Current QW landscape.} A sample of photonic discrete-time QWs to-date. Note that this table does not constitute an exhaustive list of all discrete-time photonic QWs. All experiments performed with an attenuated laser are considered to be using a 1 photon source. Note that some references quantify the quality of their results by the distance, $D = \frac{1}{2}\sum_i | P^{\text{exp}}_i - P^{\text{th}}_i |$, between the experimental, $P^{\text{exp}}_i$, and theoretical, $P^{\text{th}}_i$, probability distributions, rather than fidelity.}
  \label{tab:1}
\end{table*}

\clearpage


\providecommand{\noopsort}[1]{}

\end{document}